---

# Ideas are Not Replicators but Minds Are

***Liane Gabora***
*Department of Psychology, University of California, Berkeley*
*3210 Tolman Hall, Berkeley CA, 94720-1650 USA*
*and Center Leo Apostel for Interdisciplinary Studies (CLEA), Free University of Brussels (VUB)*
*Email: myfirstname @ berkeley.edu*
*Homepage: http://www.vub.ac.be/CLEA/liane*

**Abstract.** An idea is not a replicator because it does not consist of coded self-assembly instructions. It may *retain* structure as it passes from one individual to another, but does not *replicate* it. The cultural replicator is not an idea but an associatively-structured network of them that together form an internal model of the world, or worldview. A worldview is a primitive, uncoded replicator, like the autocatalytic sets of polymers widely believed to be the earliest form of life. Primitive replicators generate self-similar structure, but because the process happens in a piecemeal manner, through bottom-up interactions rather than a top-down code, they replicate with low fidelity, and acquired characteristics are inherited. Just as polymers catalyze reactions that generate other polymers, the retrieval of an item from memory can in turn trigger other items, thus cross-linking memories, ideas, and concepts into an integrated conceptual structure. Worldviews evolve idea by idea, largely through social exchange. An idea participates in the evolution of culture by revealing certain aspects of the worldview that generated it, thereby affecting the worldviews of those exposed to it. If an idea influences seemingly unrelated fields this does not mean that separate cultural lineages are contaminating one another, because it is worldviews, not ideas, that are the basic unit of cultural evolution.

## 1. Does Culture Evolve like Biological Lineages Do?

It is clear that cultural entities (such as ideas, habits, mannerisms, attitudes, and languages, as well as artifacts such as tools and art) evolve in the general sense of incremental change reflecting the constraints and affordances of an environment (Campbell 1987; Csanyi 1989; Cziko 1997; Gabora 1997; Hull 1988a; Lorenz 1971; Lumsden and Wilson 1981; Plotkin 1997; Popper 1963; Hofbauer and Sigmund 1988). Ideally we could flesh out a theoretical framework for this process that unifies the psychological and social sciences as did the theory of evolution for the biological sciences. Accordingly, there have been attempts to develop formal mathematical (Boyd and Richerson 1985; Cavalli-Sforza and Feldman 1981; Schuster and Sigmund 1983) and computational (Gabora 1995; Spector and Luke 1996a, b; Baldassarre 2001) models of cultural evolution. Evolutionary theory has been also applied in general terms to units of culture referred to as memes (Aunger 2000; Blackmore 1999, 2000; Dawkins 1976; Dennett 1995; Gabora 1996) as well as to the analysis of growth and change in economics (Borgers 1997; Borkar *et. al*. 1998; Metcalfe 2001; Rivkin 2001; Saviotti and Mani 1995; Witt 1992), financial markets (Farmer and Lo 1999), social customs (Durham 1991; Marsden 2001), art (Sims 1991), and the design of artifacts in primitive societies (Lake 1998). However, the endeavor to frame culture in evolutionary terms has not taken hold. It certainly hasn't had the effect of uniting previously disparate phenomena and paving the way for further inquiry, the way Darwin's theory of how life evolves through natural selection did for biology. Why not?

There are many complex aspects to this issue. This paper addressed the question of whether anything in culture, or the cognitive machinery that underlies it, constitutes a replicator, and explores implications for how to proceed with a theory of how culture evolves.

## 2. Two Kinds of Replicators

Dawkins (1976) defines a replicator as a physical entity with the following properties:

- *Longevity*-it survives long enough to self-replicate.
- *Fecundity*-at least one version of it makes multiple copies of itself.
- *Fidelity*-even after it has undergone several generations of replication, it is still almost identical to the original.

Thus, for Dawkins, a replicator makes copies of itself that survive long enough to copy *themselves*, and this continues for generations, producing lineages. The physical body that houses the replicator and gets it from place to place is referred to as the *vehicle,* or *interactor* (Hull 1988b). So long as replication is imperfect and thus introduces *variation*, and environmental interaction *selects* out the least fit, lineages evolve. But the replicator concept, as it was formulated, leaves open the crucial question: what sort of structure *can* make copies of itself? Or, to put it in more passive terms as sagely advocated by Sober and Wilson (1994), Godfrey-Smith (2000) [1] and others: what kind of structure has copies of itself made? At this point it is perhaps a good idea to note that, following in the footsteps of von Neumann (1966), when I use the word 'self' in phrases such as 'self-description' or 'self-replication', I mean not *by* the self but *of* the self.

In this section we will examine two kinds of replicators. First we consider replicators that use a self-assembly code such as the genetic code. Second we consider a more primitive sort of replicator, such as life prior to the evolution of the genetic code. In the section that follows, we examine whether cultural entities are replicators of either sort.

## 2.1 Coded Replicators

Von Neumann (1966) postulated that a genuine self-replicating system consists of coded information that can and does get used in two distinct ways. One way is as merely a description of itself, or *self-description*, that is *passively copied* to the next replicant. In this case, the code is said to be used as *uninterpreted information.* The other way is as a set of instructions for how to put together a copy of itself; that is, as *self-assembly instructions* that are *actively deciphered* to build the new replicant. In this case, the code is said to be used as *interpreted information.* To put it more loosely, the interpreting process can be thought of as 'now we make a body', and the uninterpreted use of the code as 'now we make something that can *itself* make a body'. Since biology is the field that inspired this distinction, naturally it applies here. The DNA self-assembly code is copied-without interpretation-to produce new strands of identical DNA during the process of meiosis. In successful gametes, these strands of DNA are decoded-that is, interpreted-to synthesize the proteins necessary to construct a body during the process of development.

## 2.2 Primitive Replicators

One can now ask if some component of culture is a coded replicator; that is, does von Neumann's distinction between interpreted and uninterpreted information also apply to culture? Before answering this question, however, it is instructive to see whether it applies to very primitive life-living things prior to the genetic code. Primitive life may have exhibited longevity and fecundity, but since there was not yet a genetic code for it to follow, it probably did not necessarily exhibit high fidelity. In Kauffman's (1993) autocatalytic model of the origin of life, for example, replication of the earliest lifeforms took place not according to a plan, but through happenstance interactions of catalytic polymers. Polymers catalyze reactions that generate other polymers, increasing their joint complexity, until together as a whole they form something that can more or less replicate itself. The reason this works is because when polymers interact, the number of different polymers increases exponentially, but the number of reactions by which they can interconvert increases faster than their total number. Thus, as their diversity increases, so does the probability that some subset of the total reaches a critical point where there is a catalytic pathway to every member. Such a set is *autocatalytically closed* because although none of the polymers can catalyze its own replication, each polymer can catalyze the replication of some other polymer in the set, and likewise, its own replication is catalyzed by some other member of the set. The resulting structure can be said to be self-organized because its formation was not directed by coded instructions, top-down, but rather emerged through local interactions, bottom-up. Experimental evidence for the theory with real chemistries (Lee *et al*. 1996, 1997; Severin *et al*. 1997) and computer simulations (Farmer *et al*. 1986) have been unequivocally supportive.

How does an autocatalytically closed set of polymers self-replicate? It is commonly believed that the primitive self-replicating set was enclosed in a small volume (such as a coascervate or liposome) to permit the necessary concentration of reactions (Oparin 1971; Morowitz 1992; Cemin and Smolin 1997). Since each polymer is getting duplicated somewhere in the set, eventually multiple copies of all polymers exist. The abundance of new polymers exerts pressure on the vesicle walls. This often causes such vesicles to engage in *budding*, where part pinches off and the vesicle divides in two. So long as each contains at least one copy of each kind of polymer, the set can continue to self-replicate indefinitely. Replication is far from perfect, so an 'offspring' is unlikely to be identical to its 'parent'. Different chance encounters of polymers, or differences in their relative concentrations, or the appearance of new 'food' polymers, could all result in different catalysts catalyzing a given reaction, which in turn alters the set of reactions to be catalyzed. So there is plenty of room for heritable variation.

Kauffman goes on to describe how, given say an autocatalytic set of RNA-like polymers, a phenotype-genotype distinction could come about. Some 'offspring' might have a tendency to attach small molecules such as amino acids (the building blocks from which proteins are made) to their surfaces. Some of these attachments inhibit replication and are selected against, while others favor it and are selected for. That is, we have our first indication of a division of labour between the part of the organism that interacts with the environment (the proteins), and the part concerned with replication (in this case, an RNA-based code). The advent of this code is significant for several reasons. First, the processes involved in replication and development can now be carried out recursively, and with greater precision, ensuring that structure is more or less preserved from one generation to the next. A second, related point is that by aiding the preservation of structure, it aids in the preservation of the *self-assembly capacity* of that structure, thus furthering the likelihood of continuation of the lineage beyond the next generation. Third, once the process of generating the seed of a new offspring got separated from the process of developing it into a full-fledged offspring, acquired characteristics could no longer be passed on to the next generation. Prior to coded replication, there was nothing to prohibit the inheritance of acquired characteristics. A change to one polymer would still be present after budding occurs, and this could cause other changes that have a significant effect on the lineage further downstream.

Thus, early life lacked an explicitly coded set of instructions for how to self-replicate, yet according to Kauffman's widely-accepted scenario, collectively its parts had all the instructions necessary. It consisted, in this more implicit sense, of a self-assembly code, but the two uses of it-as a self-description to be copied (without interpretation) and as (interpreted) instructions for self-replication-were neither physically nor functionally separate; the distinction between them was not so sharp as von Neumann proposed. In the very process of interpreting itself, a new replicant gets made. We will call this sort of replicator a *primitive replicator*. Whether primitive replicators are considered full-fledged replicators or not depends on whether one insists that replicators replicate with high fidelity. Since self-assembly arises not via coded instructions but through happenstance molecular interactions, there is plenty of room for error. However, since structure is preserved over generations, one can argue that they may be categorized as replicators nevertheless. Indeed, full-fledged biological replicators owe their existence to the prior existence of such primitive replicators.

## 3. Does Anything in Culture Constitute a Replicator?

We have distinguished two kinds of replicators. The first is coded replicators, which replicate using a self-assembly code, DNA-based life being the archetypal example. The second is primitive autocatalytic systems that replicate without a self-assembly code, pre-RNA life being the archetypal example. Now we ask: does anything in culture constitute a replicator of either sort?

### 3.1 Ideas and Artifacts are not Coded Replicators

First we consider whether a cultural entity constitutes a coded replicator. Thus we must determine whether it contains a self-assembly code that is used in two distinct ways: as uninterpreted information to create the seed of a new replicant, and interpreted to develop that seed into a full-fledged replicant.

#### 3.1.1 Are Cultural Entities *Interpreted?*

Let us consider first whether cultural entities are used as interpreted information, as instructions that are actively decoded to build a replicant. A transmitted idea *does* undergo something akin to interpretation when it is transmitted to and contemplated or reflected upon by a new individual. By redescribing (Karmiloff-Smith 1990, 1992) the idea in terms of what is already known, it gets more firmly rooted in the network of understandings that constitute one's internal model of reality, which for simplicity I will refer to as a *worldview*. One could say the idea builds a 'body of knowledge', a surrounding conceptual structure of related and more or less consistent ideas. The process is bi-directional; the idea not only transforms to *assimilate* (Piaget 1952) the relevant prior conceptual structure but also the worldview transforms to *accommodate* the idea (Piaget 1952). Accommodation of the idea in turn affects its further assimilation. Thus, through an assimilation/accommodation processes, cultural entities do undergo something akin to interpretation.

#### 3.1.2 Are Cultural Entities *Copied* (without Interpretation)?

Lake (1998) makes a distinction reminiscent of that made by von Neumann when he differentiates between the *expression* versus the symbolically coded *representation* of cultural information. Whereas, for example, singing a song is an expression of a musical concept, a musical score is a representation of it. As another example, the spontaneous verbal explanation of an idea is an expression, whereas the text version of it is a representation. Lake comments that some cultural entities, such as village plans, constitute *both* a representation of a symbolic plan, and an expression of that plan, because they are both expressed by and transmitted through the same material form. However neither expression nor representation is equivalent to the interpretation nor the (uninterpreted) copying of a *self-assembly code*. A village plan does not, on its own, produce little copies of itself. A musical score does not generate 'offspring scores'. The perpetuation of structure and the presence of a symbolic code do not guarantee the presence of a replicator. Symbolic coding is not enough; it must be a coded representation of the *self*.

There is in fact no reason to believe that a cultural entity such as an idea or artifact contains or consists of self-replication instructions. It is not just that it relies on the machinery of our brains to remember and, when appropriate, express or embody them. Biological replication similarly relies on the presence of certain environmental conditions in order to function properly. And it is not just that, as Sperber (1996) points out, cultural entities transform during transmission [2]. The problem is even more acute: in a cultural entity such as an idea or artifact, a self-assembly code is simply *not present*. It is one thing for an entity to continue to exist, and undergo transformations as it moves through time and space, and leave imprints on the various physical media it encounters along the way. It is quite something else for an entity to explicitly contain and interpret instructions for how to make copies of itself. The process of meiosis by which this happens in biology is extremely complex, and precisely orchestrated, according to instructions in the DNA. Chromosomes line up in pairs along the equator of a cell where their genetic information replicates, and recombines. They then pull apart on tightrope-like spindles such that one copy of each ends up in each of four daughter cells. Cultural transmission is more akin the transmission of a radio signal and its reception by one or more radios. Neither an idea nor a radio signal self-replicates in the biological sense, copying and interpreting an explicit self-assembly code. A cultural entity may *retain* structure (much as does a radio signal, or even a billiard ball as it moves from one location to another). But since no self-assembly code is present, then clearly no self-assembly code is used in the sense von Neumann identified as uninterpreted information, getting passively copied to the next replicant.

### 3.2 Interconnected Worldview as Primitive Replicator

Does this mean necessarily that *nothing* in culture constitutes a replicator? Might some kind of self-organized *network* of cultural entities constitute, if not a full-fledged coded replicator, perhaps a replicator of the primitive sort as in the case of pre-RNA life? The answer is yes, so long as in the mind there exists a set of ideas for which, for any one idea, there is an associative pathway through which it can be remembered, reconstrued, or re-described in terms of others. In other words, although ideas do not constitute replicators, interconnected *networks* of them-worldviews-*do,* of the same clumsy sort as primitive life. Just as polymers catalyze the formation of other polymers, memories and concepts trigger reminding events that evoke other memories or concepts, and this can happen recursively to generate a stream of thought. We explain a word using other words, convey an idea in terms of other ideas, understand a situation by comparing it to others.

As with the origin of life, one can consider how such a relationally structured web replicates. An adult shares concepts, ideas, attitudes, stories, and experiences with children (and other adults), influencing little by little the formation of other worldviews. Each worldview takes shape through the influence of *many* others, though some, such as those of parents and teachers, will predominate. The children expose fragments of what was originally the adult's worldview to different experiences, different bodily constraints, and thus sculpt unique internal models of the relation of self to world. Over the course of childhood, situations arise that sooner or later provide exposure to the most formative and useful knowledge, beliefs, and customs of one's society. (We all end up knowing how to read and tell the time, and having concepts of 'depth' and 'beauty' even though the particular situations that gave rise to this knowledge might differ widely from one person to another.)

It is largely the presence of abstract concepts that 'glue' different facets of the worldview together, enabling us think in relational terms. For example, the concept 'opposite' might be learned in the context of mom saying 'day is the opposite of night' and later applied as in 'black is the opposite of white'. The concept 'deep' might first be learned in the context of 'deep blue ocean' and subsequently new shades of meaning added as the child learns of 'deep in love', 'deep-fried zucchini', and 'deeply moving book'. An abstract concept spans domains that might seemingly have nothing to do with one another. It functions like a bridge that connects islands A and B in the context of someone on A needing to get to B, and then spin on its axis to connect islands C and D in the context of someone needing to get from C to D. The capacity to form concepts and engage in abstract thought is widely thought to be the most striking and distinctively human aspect of our cognition [3].

Thus, as concepts, stories, actions, and so forth get assimilated, elements of the worldviews of parents and other influential members of ones' culture get fitted together in a somewhat (but not altogether) new way. The child's worldview takes shape through the process of relating these bits and pieces to one another to form an integrated model of reality.

### 3.2.1 Conceptual Closure

It turns out that the status of a worldview as a primitive replicator may also owe to its having undergone a process of autocatalytic closure. Kauffman's solution to the question of how life began adapts readily to yield a tentative solution to the question of how culture began (Gabora 1998, submitted) and some psychological and philosophical aspects of this proposal have been explored (Gabora 1999, 2000a, 2000b, 2002). To explain how this would work, we look briefly at some aspects of the architecture of cognition. According to the doctrine of *neural re-entrance*, the same memory locations get used and reused again and again (Edelman 1987; Sporns *et. al.* 1989, 1991; Tonini *et. al*. 1992). Each of these memory locations is sensitive to a broad range of *subsymbolic microfeatures* (Smolensky 1988), or values of them (*e.g*., Churchland and Sejnowski 1992). Thus location *A* may respond preferentially to lines of a certain angle (say 90 degrees), neighboring location *B* respond preferentially to lines of a slightly different angle (say 91 degrees), and so forth. However, although location *A* responds *maximally* to lines of 90 degrees, it responds to a lesser degree to lines of 91 degrees. This kind of organization is referred to as *coarse coding*. The upshot is that storage of an item is *distributed* across a cell assembly that contains many locations, and likewise, each location participates in the storage of many items (Hinton *et al*. 1986; Palm 1980). Another result of this organization is that items stored in overlapping regions are correlated, or share features. Therefore memory is *content addressable*; there is a systematic relationship between the state of an input and the place it gets stored. This is why episodes stored in memory can thereafter be evoked by stimuli that are similar or 'resonant' (Hebb 1949; Marr 1969).

The idea of conceptual closure then is that the capacity for abstract thought arose through the onset of a tendency toward coarser coding; that is, more widely distributed storage and retrieval of memories. (This change presumably had a genetic basis and was selected for.) Thus more memory locations both (1) participate in the etching of an experience to memory, and (2) provide ingredients for the *next* instant of experience. Given that the region stored to and searched from at any given instant is wider, and because memory is content addressable, similar items are stored in overlapping regions of conceptual space, and sometimes get retrieved simultaneously. Much as catalysis increases the number of different polymers, which in turn increases the frequency of catalysis, reminding events increase the density of the stored items by triggering the emergence of abstractions (*e.g*. concepts like 'cat', 'container', or 'democracy'). Abstractions increase the frequency of reminding events because, via associative pathways, they unite all their instances (*e. g*. specific experiences of cats). Reminding events *themselves* begin to evoke reminding events recursively, thus generating streams of associative thought, which increase in both duration and frequency. In the course of these streams of thought yet more abstractions emerge, which themselves become connected in conceptual space through higher-level abstractions. Just as catalytic polymers undergo a phase transition to a state where there is a catalytic pathway to each polymer present and together they constitute a self-replicating set, memories and concepts undergo a phase transition to a state where each memory and abstraction is retrievable through a pathway of remindings/associations. Together they now constitute an autocatalytically closed, relationally structured conceptual architecture, or worldview, that both creates, and is created by, streams of thought. Thus the kind of cognitive architecture capable of sustaining cultural evolution may, like biological evolution, have originated in a phase transition to a self-organized network of catalytic relations between patterns. Kauffman (1999) suggests that this interconnected web structure is also reflected externally in the web of technological goods and services we generate.

In Farmer *et al.*'s (1987) simulation of how life could have arisen through autocatalytic closure, the probability of autocatalysis could be increased by raising either the probability of catalysis, or the complexity of the original polymers. Something similar happens here. Let us say that in the mind of a particular child, storage to and retrieval from memory are not so widely distributed. Thus the probability that one experience evokes a reminding or retrieval event is low. It is possible to make up for this by exposing the child to just the right experiences to trigger remindings, and to continue doing this until abstractions emerge, and eventually, an interconnected worldview takes shape. Individuals whose memories are not distributed widely enough to *ever* achieve worldview closure are at a reproductive disadvantage, and, over time, eliminated from the population.

Whether or not one accepts that an interconnected worldview emerges through a process of conceptual closure, it cannot be disputed that it does somehow or other emerges. Again, by interconnected I mean it has an *associative* or *relational structure*; we can define one element in terms of others, predict how a change to one element will affect another, and so forth. Thus the bottom line is that although cultural entities such as ideas and artifacts do not constitute replicators, if cultural evolution is viewed over a longer timeframe, interconnected networks of them-worldviews-can constitute cultural replicators, of the same uncoded, self-organized, primitive sort as very early life.

## 4. Implications for the Evolution of Culture

In the literature on cultural evolution it has been assumed that the basic units of cultural evolution are ideas, memes, or things like tools, fashions, languages, and so forth. Some have said that what is evolving is just the mental representations that give rise to more concrete cultural forms. The primary consequence of the line of argument presented in the preceding sections is that the level at which cultural self-replication is taking place, and the basic unit of cultural evolution, is the entire interwoven conceptual structure of the mind; the worldview. Moreover, the worldview is not a genuine coded replicator, but a primitive, uncoded replicator, more like pre-RNA than present-day life in its mode of evolving. These conclusions may appear strange at first glance, so let us now examine them, and their evolutionary significance, in more detail.

### 4.1 What Evolves is Worldviews, Not Ideas

The skeptic might argue that even if ideas are not strictly replicators, we can treat them as a basic unit of cultural evolution nevertheless, using the following argument from biology. Individual genes are not replicators; alone they cannot self-replicate (Lewontin 1991; Nanay 2002). But since (until the recent advent of genetic engineering) they always come packaged together as genomes in the cellular milieu of an organism that supports their replication (the vehicle or interactor), they can be treated as such. In other words, although strictly speaking they are not replicators, they exhibit replicator dynamics because they exist only in the context of entities that *do* constitute replicators. A gene for eye color is always found packaged with a gene for blood type, as well as all the other genes necessary to build an organism.

Is this also the case for ideas? Might they *also* exhibit replicator dynamics because they are part of a package-an interconnected network of ideas-that does constitute a replicator? The problem is that cultural entities do not replicate together as genes do; that is, they do not get transmitted as a package containing some version of all other cultural entities. Only a fragment of the vast interwoven network of knowledge, stories, attitudes and beliefs that constitute ones' internal model of the world, or worldview, gets culturally expressed at any one time. You may explain with both words and gesture how to change a tire. But when you explain how to change a tire, you do not simultaneously tell your favorite ghost story, vent your emotions playing a piano concerto, and express every other bit of knowledge that lies latent within you.

There is, however, a way out of this problem. It can be resolved by clumping all cultural entities together as manifestations of the worldviews that generate them. Thus a gesture or idea is how a worldview reveals itself in a particular context. When you explain how to change a tire, certain facets of your worldview are revealed, while playing a piano concerto reveals others. The situation of needing to explain how to change a tire simply 'sliced through' your worldview in such a way that some parts of it were not in evidence. A painting is viewed not just as a particular design created with particular colours of paint, but as evidence concerning the state of the worldview that generated it. Thus we do not have to worry that paintings themselves, nor the mental representations responsible for them, do not constitute genuine replicators. The painting plays its role in the evolution of culture by revealing some aspect of the artist's worldview (which *is* a replicator) and thereby affecting the worldviews (other replicators) of those who admire it.

Note that Sterelny *et al*. (1996) argue convincingly that replicators are selected for, not just their capacity to get themselves replicated, but also for their contribution to their own development. If one considers ideas to be replicators, it is hard to see how they could be selected on the basis of their contribution to their own further development without bringing in the notion of a network of related ideas. But if the worldview is the replicator, and the idea merely a reflection of the current state of the worldview, there is no longer a problem. What is *developing* is an internal model of how the various aspects of the world relate to one another. Parental worldviews composed of ideas, attitudes, and so forth that foster the development of a more or less coherent, useful, and satisfying worldview in the child would seem to be at a selective advantage.

### 4.2 Evolving without Copying from a Code

Sperber (1996) argues that a child's views may be acquired through a process of not copying but influencing, on many occasions and from many sources, or through "convergence... toward some psychologically attractive type of views in the vast range of possible views" (p. 106). He argues that, although no copying is taking place, cultural entities (or representations, as he refers to them) evolve nevertheless.

Some might find this unconvincing. Elements of the physical world influence one another and converge toward stable attractor states all the time; why should this give you *evolution* ? But Sperber's position that culture evolves without copying is strengthened considerably when we consider that what is evolving is not separate ideas or attitudes but worldviews, *i.e.* interrelated networks of them. Worldviews are primitive replicators, which replicate without copying from a code. Underlying any idea is a web of assumptions that render it not only possible to be expressed, but sensible and worth considering in the first place. Even when one does not believe an argument or adopt an attitude, one's worldview is nevertheless affected by (potentially indirect) exposure to this underlying web of assumptions. As Sperber points out, it is not necessary that an idea be understood or even believed to be true for its influence to spread. For example, if Ann's argument for why free trade is good appears flawed to Bill, and thus strengthens Bill's conviction that free trade is *bad*, the structure of Ann's argument has spread even if the final conclusion has not [4]. Thus even if no particular overtly expressed act or belief gets explicitly copied, through exposure to parents and other influential members of society, a general framework for how the world hangs together falls into place in the child's mind, and this worldview will have some likeness to that of its predecessors. The worldview of a child is a replicant of the worldviews of its parents (and others) so long as the fidelity requirement is not applied stringently.

Again, this has some similarity to the situation of the evolution of life prior to the genetic code. Since there was no code to copy from, there was no explicit copying going on. The presence of a given catalytic polymer, say polymer X, simply speeded up, that is, influenced, the rate at which certain reactions took place, while another catalytic polymer, say Y, influenced the reaction that generated X. Of course, once the genetic code came about, self-replication became more explicitly a process of copying from a single source (that is, only one parent was involved in any one copying event). But with the advent of sexual reproduction this again became less the case; an offspring was now constructed from more than one parent (generally two). The bottom line is that even in biological evolution, self-replication is far from always a process of copying from a single source; it can take place also through multiple sources, without following a code.

### 4.3 Inheritance of Acquired Characteristics

We saw that, like autocatalytic sets of polymers, worldviews do not evolve with the efficiency of a symbolic code, but in a piecemeal manner, largely through the expression, assimilation, and accommodation of cultural entities in the course of social interaction. This has important consequences for how traits get passed on from one generation to the next.

Let us briefly re-examine one aspect of the significance of the transition from uncoded to coded replicators in biology. With the advent of the genetic code, acquired characteristics were no longer passed on to the next generation. The presence of explicit self-assembly instructions, and the separate use of them as interpreted and uninterpreted information, meant that the replicant did not incorporate changes to the parent that occurred during development or in maturity. Thus for example, if one cuts off the tail of a mouse, its offspring will have tails of a normal length. Prior to coded replication, this was not the case. A change to one polymer would still be present in an offspring after budding occurred, and this could cause other changes that have a significant effect on the lineage further downstream. There was nothing to prohibit the inheritance of acquired characteristics.

Note that it is often said that because acquired traits are inherited in culture, culture should not be viewed in evolutionary terms. It is somewhat ironic that this critique also applies to the earliest stage of biological evolution itself. What was true of early life is also true of the replication of worldviews: there is nothing to prohibit the inheritance of acquired characteristics [5]. Consider again the example of the musical score. One can imagine a sort of molecular construction that could be understood by a musician as a musical score and played accordingly, and that had, encoded in it, instructions for how to piece together molecules in its surrounding medium to generate identical or similar molecular musical scores. One could modify the 'parent' score, and this would accordingly change how it was played. Unless one changed that particular part of the score that dealt with making copies of itself, the next generation of musical scores would not be affected [6]. Thus, changes acquired during any particular generation of this score could not be passed on. But this is not the case for the sort of musical score with which we are familiar. They do not contain a self-assembly code. (Nor do they need one, since *we* do the replicating of cultural entities, contextually modifying them according to our wishes, needs, and desires.) So once again there is nothing to prohibit the inheritance of acquired characteristics. We hear a joke and, in telling it, give it our own slant; we create a disco version of Beethoven's Fifth Symphony and a rap version of that. Moreover since it is not Beethoven's Fifth per se that is evolving, but the worldviews of individuals exposed to it, we do not have to worry about 'contamination of cultural lineages' if listening to it has an affect on, say, the book one is writing.

## 5. Conclusions

We have seen that *an* idea is not a replicator because it does not consist of coded self-assembly instructions. It does not make copies of itself; it may *retain* structure as it passes from one individual to another, but does not *replicate* it. Does this mean that culture does not constitute an evolutionary process, or that an evolutionary process does not require replicators?

Some are happy to say that evolution does not require replicators of any sort (see commentary on a recent target article by Hull *et. al*. 2001). However, in the case of culture, it is not necessary to abandon the view that replicators play a vital role in evolution if we posit the notion of a *primitive replicator*. A primitive replicator generates self-similar primitive replicators, but in a piecemeal manner, and with low fidelity. A pre-RNA autocatalytic set is an example of a primitive replicator. A worldview-an interconnected *network* of ideas that together constitute an internal model of the world-is another example. Just as polymers catalyze reactions that result in the formation of other polymers, the retrieval of an episode or concept (or a creatively reconstructed blend of *many* items) from memory can in turn trigger other episodes or concepts, which get recursively re-described in terms of one another in a stream of thought. Each episode or concept thereby gets integrated into this associative structure, which evolves idea by idea, largely through social exchange, sometimes mediated by artifacts. As worldviews become increasingly complex, the artifacts they manifest in the world become increasingly complex, which necessitates even more complex worldviews, *et cetera*. Through this evolution of worldviews, today's culture is rooted in the culture of the past. Because the process of worldview formation works through 'bottom-up' interactions rather than a 'top-down' code, characteristics accumulated in one generation can be transmitted to the next.

In dis-analogy with the gene, an idea cannot be assumed to act like a replicator because it does not come packaged with the other elements of one's worldview. However, this problem is bypassed by viewing any idea as evidence concerning the state of the worldview that generated it. The idea participates in the evolution of culture by revealing some aspect of this worldview (which *is* a replicator) and thereby affecting the worldviews (other replicators) of those exposed to it. And if it is worldviews that are evolving, not ideas, then if this idea eventually exerts an effect on seemingly unrelated fields this does not mean that separate cultural lineages are contaminating one another.

In summary then, culture may be viewed as a process of evolution, but the replicator is not a cultural entity such as an idea, attitude, or piece of knowledge. It is an associatively-structured, interconnected network of them; that is, an internal model of the world, or worldview. Moreover, if we wish to describe culture in evolutionary terms we must be willing to forgo the requirement that its evolution involve coded replicators, and that they replicate with high fidelity. A worldview is a replicator of a primitive, uncoded sort, more like that of the earliest form of life than present-day life, and subject to the inheritance of acquired characteristics.

## Acknowledgements

I would like to thank Kim Sterelny and an anonymous referee for helpful comments. I would also like to acknowledge the support of Grant G.0339.02 of the Flemish Fund for Scientific Research.

## Footnotes

[1] Godfrey-Smith (2000) suggests that we use the notion of *replicate* to indicate more explicitly that replication involves the transmission of both *resemblance* and *causal relations*, where replicate is defined as follows: "Y is a replicate of X if and only if: (i) X and Y are similar (in some relevant respects), and (ii) X was causally involved in the production of Y in a way responsible for the similarity of Y to X." Although I am in agreement with Godfrey-Smith about the transmission of resemblance and causal relations, I find it more useful to use the word 'replicate' as a verb, *i.e*. to refer to what a replicator does.

[2] Sperber argues that because cultural entities transform (not just when they are contemplated, but even *during* the transmission process), 'replication' of them can be viewed as the limiting case of null transformation. However, because cultural entities lack self-assembly instructions, even in this null case the situation is unlike that of DNA-based replication(although at a gross level the result may in some cases be similar).

[3] It is not universally thought that the capacity for concept formation, abstract thought, or creativity that is most uniquely human, and that brought about the origin of complex human culture. Some say it is the capacity to imitate (*e.g*. Blackmore 1999; Boyd & Richerson 1985, Tomasello 2000). Others have argued that it is a 'theory of mind', or understanding of the intentionality of others (*e.g*. Heyes 1998). Clearly each of these is important. My reason for emphasizing concept formation and abstract thought is that they are stepping stones toward an interconnected internal model of the world, and this relational structure could in turn aid both the capacity to imitate and to model the intentionality of others. There is in fact considerable debate over when the cognitive capacities underlying complex culture came about. Corballis (2002) suggests that as early as two million years ago we evolved an abstract generative capacity that enabled us to create new, potentially complex, recursive structures in a variety of different domains or contexts such as language, music, and the production of artifacts. Mithen (1998) suggests that our unique cognitive abilities arose as recently as 50,000 years ago, due to an enhanced capacity for abstract thought, leaving us with an ability to map, explore, and transform conceptual spaces. There is also debate concerning how abstract concepts derive their meaning. Some authors stress that a concepts' meaning comes from the web of inferential connections to other concepts (e.g. Heyes 1998), while other authors maintain that it comes from their being grounded, or having multiple relations to the external world (*e.g*. Sterelny 2000).

[4] One can say that the 'free trade is good' and 'free trade is bad' positions are *orthogonal* but nevertheless *compatible;* that is, opposite and contradictory, but revealed by the same kind of context or measurement (Gabora & Aerts 2002a, b).

[5] Whether this implies that cultural evolution is Lamarckian is a controversial matter. While

some have said that it is (*e.g*. Boyd & Richerson 1985; Dawkins 1982; Gabora 1997), Hull (1988) insists that an evolutionary process is not Lamarckian unless change is not only passed on but incorporated into the genetic material.

[6] Furthermore, unless like the genetic code this musical code was very redundant, or had some other feature that made it robust. It would likely be so highly constrained that it might well be subject to error catastrophe (Eigen & Schuster 1979). Any change to it would destroy its self-replication capacity.